\documentclass[%
 twocolumn,
 amsfonts,amsmath,amssymb,
 aps,
prd,
floatfix,
showpacs,
longbibliography
]{revtex4-2}

\usepackage{graphicx}
\usepackage{dcolumn}
\usepackage{bm}
\usepackage{wasysym}
\usepackage[dvipsnames]{xcolor}
\usepackage{aas_macros}
\definecolor{ReflexBlue}{rgb}{ .0902,.0902,.5882}
\usepackage{hyperref}
\hypersetup{
    colorlinks=true,
    linkcolor=ReflexBlue,
    filecolor=magenta,      
    urlcolor=ReflexBlue,
    citecolor=ReflexBlue,
}
\let\Phi\varPhi

\newcommand{\ee}{e}  
\newcommand{\com}{\mathcal{C}}
\newcommand{\rhoc}{\epsilon_{\text{c}}}
\newcommand{\tol}{\text{\tiny Tol}}
\newcommand{\imp}{\text{\tiny mod}}

\begin{document}

\title{New interior model of neutron stars}
\author{Camilo Posada}\email{camilo.posada@physics.slu.cz}
\affiliation{Research Centre for Theoretical Physics and Astrophysics, Institute of Physics, Silesian University in Opava, Bezru\v{c}ovo n\'{a}m. 13, CZ-746 01 Opava, Czech Republic}

\author{Jan Hlad\'ik}\email{jan.hladik@physics.slu.cz}
\affiliation{Research Centre for Theoretical Physics and Astrophysics, Institute of Physics, Silesian University in Opava, Bezru\v{c}ovo n\'{a}m. 13, CZ-746 01 Opava, Czech Republic}

\author{Zden\v{e}k Stuchl\'ik}\email{zdenek.stuchlik@physics.slu.cz}
\affiliation{Research Centre for Theoretical Physics and Astrophysics, Institute of Physics, Silesian University in Opava, Bezru\v{c}ovo n\'{a}m. 13, CZ-746 01 Opava, Czech Republic}
\date{\today}

\begin{abstract}
The Tolman VII solution is considered by some as one of the few analytical solutions to Einstein's equations, which describes approximately well the interior of neutron stars (NSs). This solution is characterized by the mass $M$, radius $R$, and an energy density that varies quadratically with the radial coordinate $r$. Recently, Jiang and Yagi proposed a modification of this solution, the so-called modified Tolman VII (MTVII) solution, by introducing an additional quartic term to the energy density radial profile. The MTVII solution is an approximate solution to Einstein's equation, which includes a new parameter $\alpha$ that allows the solution to have a better agreement with the energy density profiles for realistic NSs. Here we consider the MTVII solution, showing that for certain values of the parameter $\alpha$ and compactness $\com$ this solution manifests a region of negative pressure near the surface which leads to negative values of the tidal Love number. To alleviate these drawbacks, we introduce an exact version of the MTVII solution obtained by solving numerically Einstein's equations for the MTVII energy density profile. As an application of our new exact MTVII (EMTVII) solution, we calculate the tidal Love number and tidal deformability, as a function of $\com$, for different values of the parameter $\alpha$. We find that the EMTVII solution predicts a positive tidal Love number for the whole range of allowed values of parameters $(\com,\alpha)$, in agreement with previous results for realistic NSs.
\end{abstract}

\maketitle

\section{Introduction}\label{Intro}

The Tolman VII (TVII) solution~\cite{Tolman39}, of the Einstein gravitational field equations, has attracted some interest from the astrophysical community. There are many reasons for its popularity. First, it is a simple analytical solution of Einstein's equations describing relatively well the spacetime structure of the interior of neutron stars (NSs)~\cite{Lattimer:2000nx, Postnikov2010}. The TVII solution can be constructed as being based on a simple quadratic radial profile of the energy density, which enables exact solutions for the radial profiles of the metric components and pressure.\\
\indent The stability under radial oscillations of the TVII solution was studied by \cite{Negi:2001ApS, Moustakidis:2016ndw}, its tidal Love numbers were computed by~\cite{Postnikov2010} and some of its general properties have been further elucidated in \cite{Raghoonundun:2015wga}. On the other hand, the TVII solution allows the existence of ultracompact objects containing a region of trapped null geodesics \cite{Abramowicz:1993irb, Neary:2001ai, Stuchlik:2016xiq, Novotny:2017cep, Stuchlik:2017qiz, Stuchlik:2021coc, Stuchlik:2021EPJP}. It is relevant that the radius of the trapping TVII spheres can significantly overcome the limit $R=3M$ valid for the interior Schwarzschild spacetime, reaching a radius that could be in agreement with the estimated radius of observed NSs \cite{LIGOScientific:2018cki, De:2018uhw, Miller:2019cac, Riley:2019yda,Miller:2021qha,Riley:2021pdl}, implying thus the astrophysical plausibility of trapping effects.\\
\indent To improve the agreement of the TVII solution with the energy density radial profiles predicted by realistic equations of state (EOS), corresponding to a proper description of the high-density NS matter, Jiang and Yagi~\cite{Jiang2019} introduced the modified Tolman VII (MTVII) solution by adding a quartic term to the energy density radial profile. They introduced an additional parameter $\alpha$ to describe more accurately the realistic energy density profiles. However, it was found that for this modification not all of the Einstein equations can be solved exactly, therefore certain approximate analytical expressions were proposed keeping a strong analogy with the original TVII solution.\\
\indent Some general properties of the MTVII solution have been studied in the literature; for instance, the so-called analytic I-Love-$\com$ relations were constructed in~\cite{Jiang2020} by extending the model to slowly rotating or tidally deformed configurations. In Ref.~\cite{Posada2021} we studied the stability against radial perturbations, using Chandrasekhar's method of infinitesimal radial perturbations. We found that the MTVII model is stable for a wide range of compactness $\com$, in the allowed regime of $\alpha$. However, as we shall discuss later, we found recently some inconsistencies in the MTVII solution, particularly in the modified expression for the pressure introduced in~\cite{Jiang2019}, which for certain configurations, leads to negative values near the surface. As a consequence, for certain configurations, the MTVII model predicts negative tidal deformability.\\
\indent For this reason, we present here an exact MTVII solution (i.e., EMTVII) by exactly solving (numerically) the set of Einstein's equations for the quartic energy density model of the MTVII solution. We compare our solution with the approximate radial profiles of the $g_{tt}$ metric component and pressure of MTVII. As a further application of our new solution, we compute the tidal deformability, as a function of the compactness, for different values of the parameter $\alpha$. We found that our solution predicts positive tidal deformability, in the full allowed range of values $(\com,\alpha)$, in agreement with what is expected for realistic NSs. Finally, we compare our results of the tidal deformability for the EMTVII solution with those predicted by MTVII, and we provide some observational constraints based on GW170817.\\
\indent The paper is organized as follows. In Sec.~\ref{sec:2} we review the original TVII solution and the modified version, or MTVII. In Sec.~\ref{sec:3} we provide an analysis of the MTVII solution, with particular emphasis on the profiles for pressure. In Sec.~\ref{sec:4} we present the new EMTVII and compare its $g_{tt}$ metric component and pressure profiles with the MTVII model. As an application, in Sec.~\ref{sec:5} we present results of the tidal Love number $k_2$ for the EMTVII and MTVII solutions and compare them. Furthermore, we translate our results in terms of the dimensionless tidal deformability $\bar{\Lambda}$ to include current constraints on the tidal deformability obtained from the event GW170817. Throughout this paper we use geometrized units, $c=G=1$.   

\section{Tolman's method for the solution of a fluid in equilibrium}\label{sec:2}
Following Tolman \cite{Tolman39}, we consider a static and spherically symmetric matter distribution. The relevant line element in this case takes the standard Schwarzschild form
\begin{equation}\label{metric}
ds^2 = -\ee^{\nu(r)}dt^2+\ee^{\lambda(r)}dr^2 + r^2\left(d\theta^2 + \sin^2\theta \, d\varphi^2\right)\, ,
\end{equation}
where $\nu$ and $\lambda$ are functions of $r$. We describe the matter inside the configuration as a static perfect fluid; thus its energy-momentum tensor takes the form
\begin{equation}\label{tmunu}
T_{\mu\nu} = (\epsilon + p)u_{\mu}u_{\nu} + pg_{\mu\nu}\, ,
\end{equation}
where $\epsilon$ denotes the energy density, $p$ is the pressure, and $u^\nu = dx^\nu / d\tau$ is the four-velocity. Substituting
Eqs.~\eqref{metric} and \eqref{tmunu} into Einstein's equations $G_{\mu\nu} = 8\pi T_{\mu\nu}$, one finds~\cite{Tolman39}
\begin{multline}\label{einstein1}
    \frac{d}{dr}\left(\frac{\ee^{-\lambda}-1}{r^2}\right) + \frac{d}{dr}\left(\frac{\ee^{-\lambda}\nu^\prime}{2r}\right) \\
        + \ee^{-(\lambda+\nu)}\frac{d}{dr}\left(\frac{\ee^{\nu}\nu'}{2r}\right) = 0\, ,
\end{multline}
\begin{equation}\label{einstein2}
    \ee^{-\lambda}\left(\frac{\nu^\prime}{r}+\frac{1}{r^2}\right) - \frac{1}{r^2} = 8\pi p\, ,
\end{equation}
\begin{equation}\label{einstein3}
    \frac{dm}{dr}=4\pi r^2 \epsilon\, ,
\end{equation}
where $'\equiv d/dr$ and $m(r)$ corresponds to the mass enclosed in the radius $r$. It is conventional to relate the mass $m(r)$  and the metric element $e^{-\lambda}$ in the form
\begin{equation}\label{einstein4}
  \ee^{-\lambda(r)} \equiv 1 - \frac{2m(r)}{r}\, .
\end{equation}
This system will be closed once an EOS, connecting pressure with energy density, is provided. Tolman~\cite{Tolman39} realized that by choosing, conveniently, certain functions for $e^{\lambda}$ and $e^{\nu}$, the set of Einstein's equations~\eqref{einstein1}--\eqref{einstein3}, can be integrated analytically and in a straightforward manner. One of the solutions found by Tolman using this method is the so-called, TVII solution which we discuss in the next subsection.

\subsection{Tolman VII solution}\label{sec1-1}
In this section we discuss the exact solution to the Einstein equations known as the TVII solution~\cite{Tolman39,Lattimer:2000nx}. Here we  follow the notation used by~\cite{Jiang2019}. Tolman assumed $\ee^{-\lambda(r)}$ in the form
\begin{equation}\label{grrTol}
  \ee^{-\lambda_\tol} = 1 - \frac{8\pi}{15}\epsilon_{\text{c}}R^2 x^2 (5 - 3x^2)\, ,
\end{equation}
where $x \equiv r/R$, with $R$ denoting the radius of the star and $\epsilon_{\text{c}}$ is the central energy density (henceforth the label “Tol" will indicate quantities associated with the original TVII solution). With this ansatz, the energy density $\epsilon$, mass $m$, and pressure $p$ are found to be
\begin{equation}\label{rhoTol}
  \epsilon_\tol = \epsilon_\mathrm{c} (1 - x^2), \qquad m_\tol = \frac{M}{2} x^3 (5 - 3x^2)\, ,
\end{equation}
\begin{equation}\label{pTol}
  \frac{p_\tol}{\epsilon_\mathrm{c}} = \frac{1}{15} \left[\sqrt{\frac{12 \ee^{-\lambda_\tol}}{\com}}\tan\phi_\tol-(5 - 3x^2)\right]\, .
\end{equation}
Here $M = m(R)$ is the total mass, and $\com\equiv M/R$ is the compactness of the configuration. Note that the energy density vanishes at the stellar surface $r = R$. The $g_{tt}$ metric component results
\begin{equation}\label{gttTol}
  \ee^{\nu_\tol} = C_1^\tol \cos^2 \phi_\tol\,,
\end{equation}
where
\begin{equation}\label{phiTol}
  \phi_\tol = C_2^\tol - \frac{1}{2}\log \left(x^2 - \frac{5}{6} + \sqrt{\frac{5\ee^{-\lambda_\tol}}{8\pi \epsilon_\mathrm{c} R^2}}\right).
\end{equation}
Here $C^\tol_1$ and $C^\tol_2$ are constants of integration given by
\begin{equation}\label{C1Tol}
C_{1}^\tol=1-\frac{5\com}{3},
\end{equation}
\begin{equation}\label{C2Tol}
C_{2}^\tol=\arctan{\sqrt\frac{\com}{3(1-2\com)}}+\frac{1}{2}\log\left(\frac{1}{6}+\sqrt{\frac{1-2\com}{3\com}}\right)
\end{equation}
Let us recall certain restrictions for the physical plausibility of the TVII solution. For instance, the pressure is finite for $C < 0.3862$ and the dominant energy condition (DEC) is valid for $\com<0.3351$ \cite{Posada2021}.

\subsection{Modified Tolman VII solution}\label{sec1-2}
In this subsection we discuss the modified MTVII solution proposed by~\cite{Jiang2019}. In this new model, the energy density $\epsilon$ is assumed as a quartic function of the radial coordinate in the form
\begin{equation}\label{rhoMod}
  \epsilon_\imp = \rhoc\left[1 - \alpha x^2  + (\alpha - 1)x^4\right]\, ,
\end{equation}
where $\alpha$ is a new free parameter of the solution. This MTVII solution seems to model more accurately the energy density profile for realistic EOS of NSs, as compared with the original TVII solution~\cite{Jiang2019}. Substituting Eq.~\eqref{rhoMod} into Eqs.~\eqref{einstein3} and \eqref{einstein4} we can solve immediately  for $m$ and $e^{-\lambda}$
\begin{equation}\label{massMod}
  m_\imp = 4\pi \epsilon_\text{c} R^3 x^3 \left(\frac{1}{3}-\frac{\alpha}{5}x^2 + \frac{\alpha - 1}{7}x^4\right)\, ,
\end{equation}
\begin{equation}\label{grrMod}
  \ee^{-\lambda_\imp}= 1 - 8\pi \rhoc R^2 x^2 \left[\frac{1}{3}-\frac{\alpha}{5}x^2 + \frac{(\alpha - 1)}{7}x^4\right]\, .
\end{equation}
Given the complexity introduced by the parameter $\alpha$, in principle, it is not possible to find analytic expressions for the pressure $p$ and the $g_{tt}$ metric component. Therefore, \cite{Jiang2019} proposed the following approximate solution
\begin{equation}\label{gttMod}
  \ee^{\nu_\imp} = C_1^\imp\cos^2\phi_\imp\, ,
\end{equation}
with
\begin{equation}\label{phiMod}
  \phi_\imp = C^\imp_2 - \frac{1}{2}\log\left(x^2 - \frac{5}{6} + \sqrt{\frac{5\ee^{-\lambda_\tol}}{8\pi\rhoc R^2}}\right),
\end{equation}
where $C_{\imp}^1$ and $C_{\imp}^2$ are integration constants which read now
\begin{equation}\label{C1Mod}
C_{1}^\imp=(1-2\com)\left\{1+\frac{8\pi\rhoc R^2(10-3\alpha)^2 (15-16\pi\rhoc R^2)}{3\left[105+16\pi\rhoc R^2 (3\alpha-10)\right]^2}\right\},
\end{equation}
\begin{multline}\label{C2Mod}
C_{2}^\imp = \arctan\left[-\frac{2(10-3\alpha)\sqrt{6\pi\rhoc R^2(15-16\pi\rhoc R^2)}}{48\pi(10-3\alpha)\rhoc R^2-315}\right] \\
+\frac{1}{2}\log\left[\frac{1}{6}+\left(\frac{5}{8\pi\rhoc R^2}-\frac{2}{3}\right)^{1/2}\right].
\end{multline}
With these expressions, the pressure can be obtained directly from Eq. \eqref{einstein2} as
\begin{equation}\label{pOrig}
  \tilde{p}_\imp(r) = \frac{1}{8\pi}\left[\ee^{-\lambda_\imp}\left(\frac{\nu^\prime_\imp}{r}+\frac{1}{r^2}\right)-\frac{1}{r^2}\right]\,.
\end{equation}
However,~\cite{Jiang2019} found that Eq.~\eqref{pOrig} gives a central pressure which is off by roughly $20\%$ from the numerical results. Moreover, it gives an unphysical negative pressure region near the surface of the star. In order to fix these drawbacks, \cite{Jiang2019} introduced a corrected expression given by
\begin{multline}\label{pMod}
  p_\mathrm{mod} = \epsilon_\text{c} \left[\left(\frac{\ee^{-\lambda_\text{Tol}}}{10\pi \epsilon_\text{c}R^2}\right)^{1/2}\tan \phi_\text{mod} +\frac{1}{15}(3x^2 - 5)+\right. \\ \left.\frac{6(1-\alpha)}{16\pi \epsilon_\text{c} R^2(10 - 3\alpha) - 105}\right]\,.
\end{multline} 
From Eqs. \eqref{rhoMod}--\eqref{pMod}, it can be observed that the MTVII solution reduces to the original TVII when $\alpha=1$. 
The MTVII solution is determined by four parameters, namely, $(\epsilon_\text{c}, \mathcal{C}, R, \alpha)$. However, Eq.~\eqref{massMod} gives the condition for the total mass $M = m_\mathrm{mod}(1)$, such that
\begin{equation}\label{comMod}
  8\pi \epsilon_\mathrm{c} R^2 = \frac{105\mathcal{C}}{10-3\alpha}\,.
  \end{equation}
Thus, the MTVII solution is fully determined by three parameters $(\epsilon_\text{c}, \mathcal{C}, \alpha)$. We will use this convention in the subsequent calculations.    

\section{Analysis of the MTVII solution}\label{sec:3}

Let us analyze some general properties of the MTVII solution. We consider the parameter $\alpha$ in the range $\alpha\in[0, 2]$. This range is restricted by the solution as follows: for $\alpha < 0$ the density is a nonmonotonically function of $r$ which is not consistent with the realistic EOS for NSs. On the other hand, for $\alpha > 2$ the energy density becomes negative, which is unphysical. Jiang and Yagi~\cite{Jiang2020} restricted their analysis to $\alpha \in [0.4, 1.4]$ with $\com\in[0.05,0.35]$, which seems to be the relevant range for comparison to realistic EOS for NSs. However, we consider that this range is too restrictive in the sense that one could find a NS model which lies outside that regime, therefore it is advantageous to consider the whole allowed range $\alpha\in[0, 2]$.

\begin{figure*}
    \includegraphics[width=.44\linewidth]{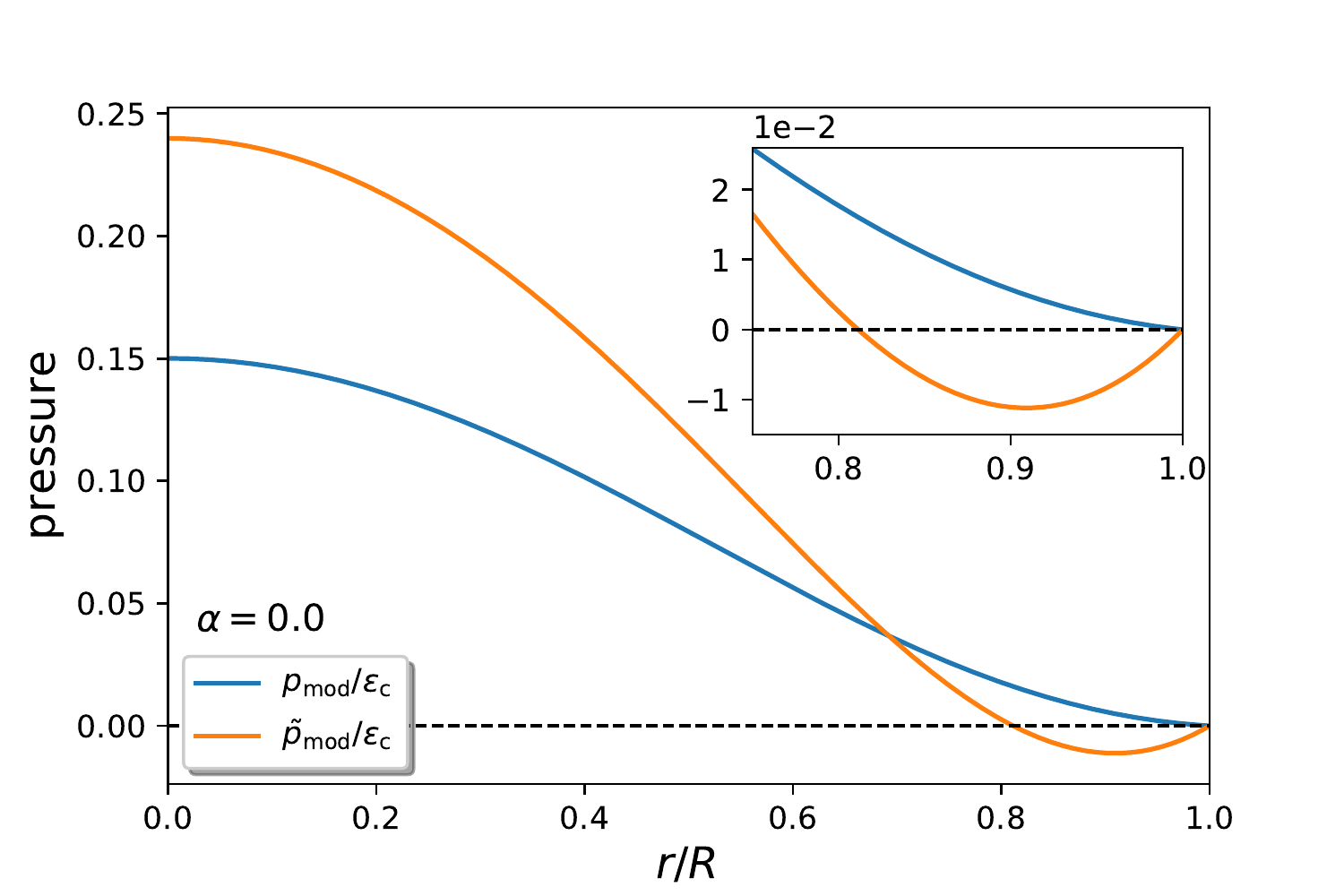}\hfill
    \includegraphics[width=.44\linewidth]{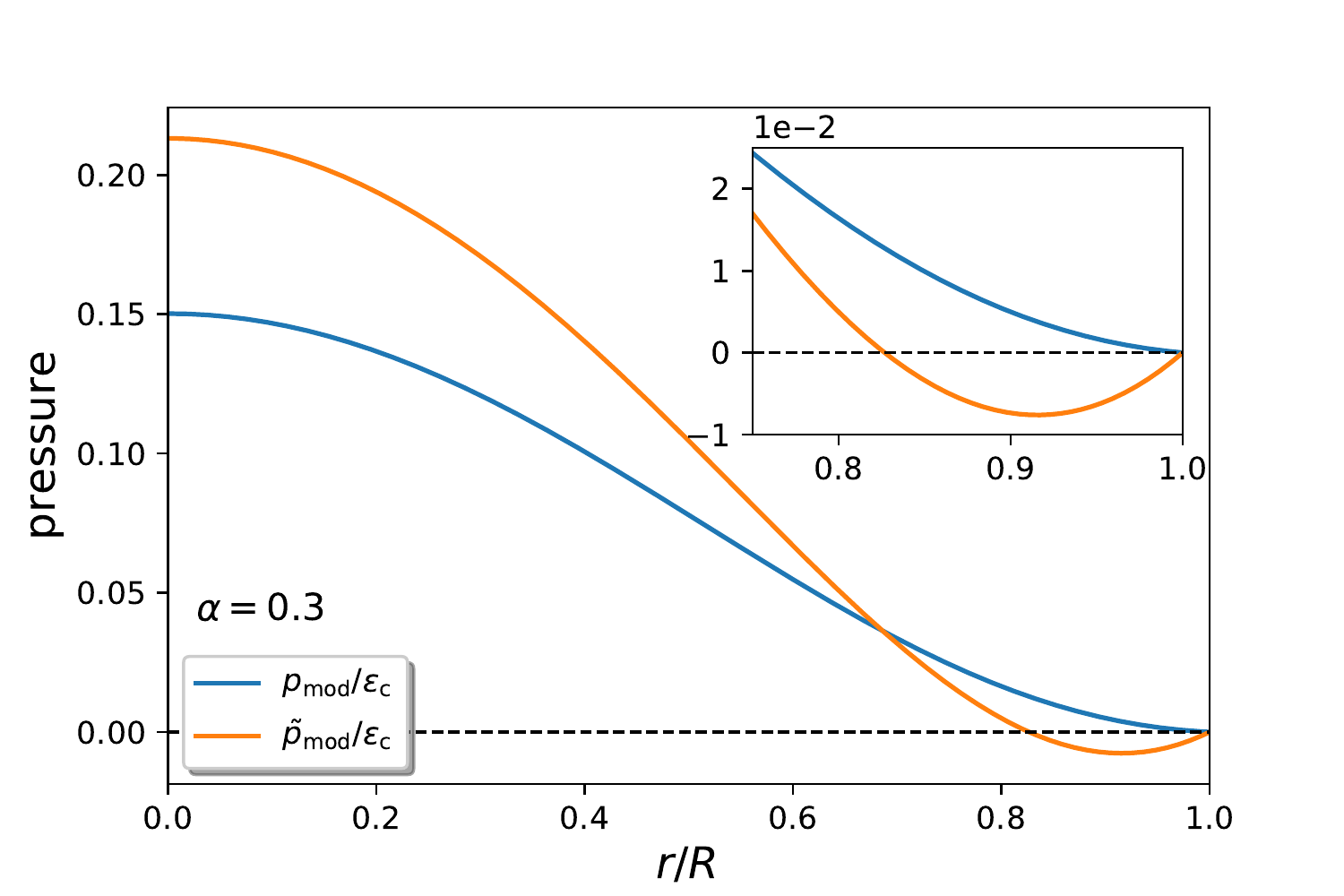}
    \includegraphics[width=.44\linewidth]{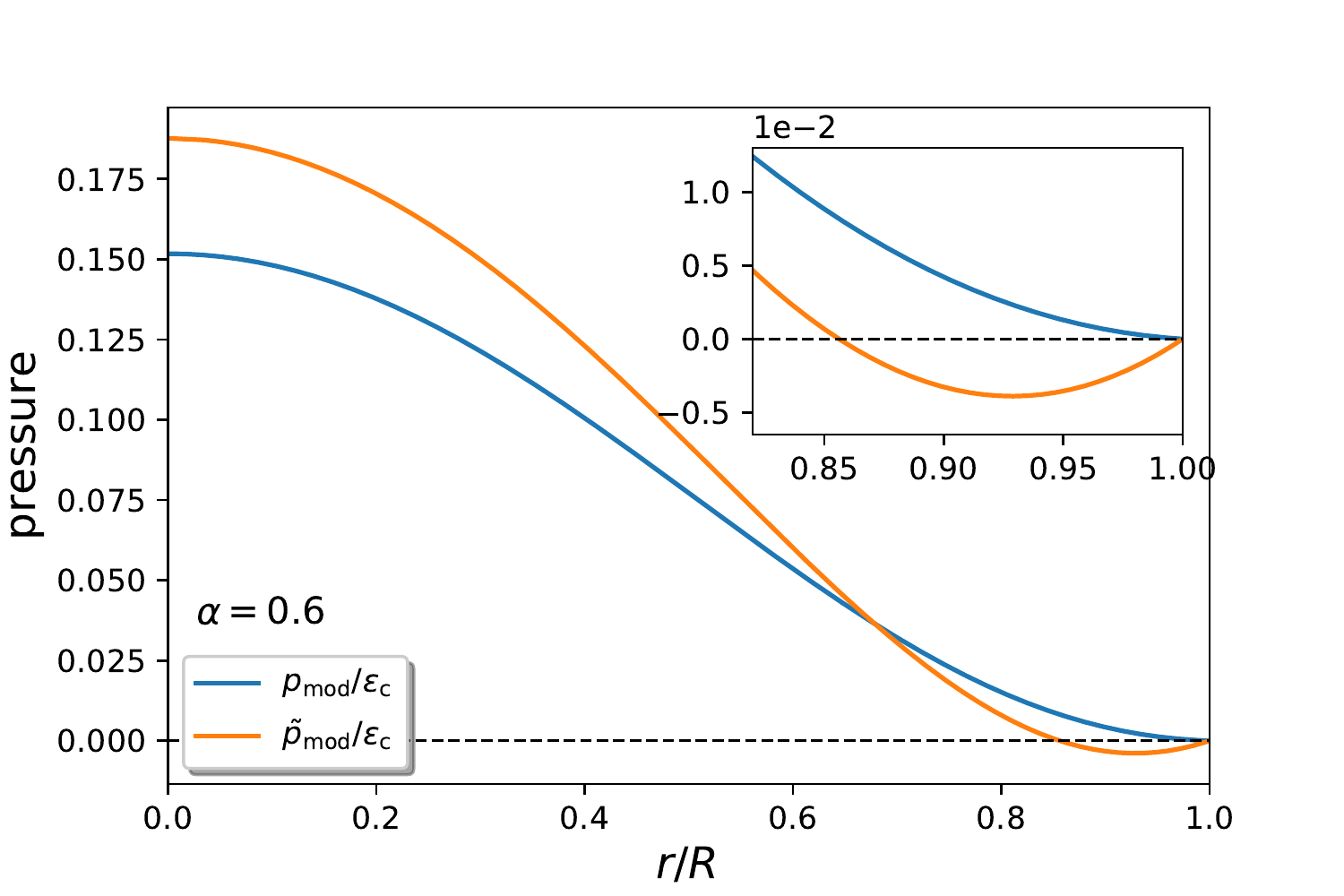}\hfill
    \includegraphics[width=.44\linewidth]{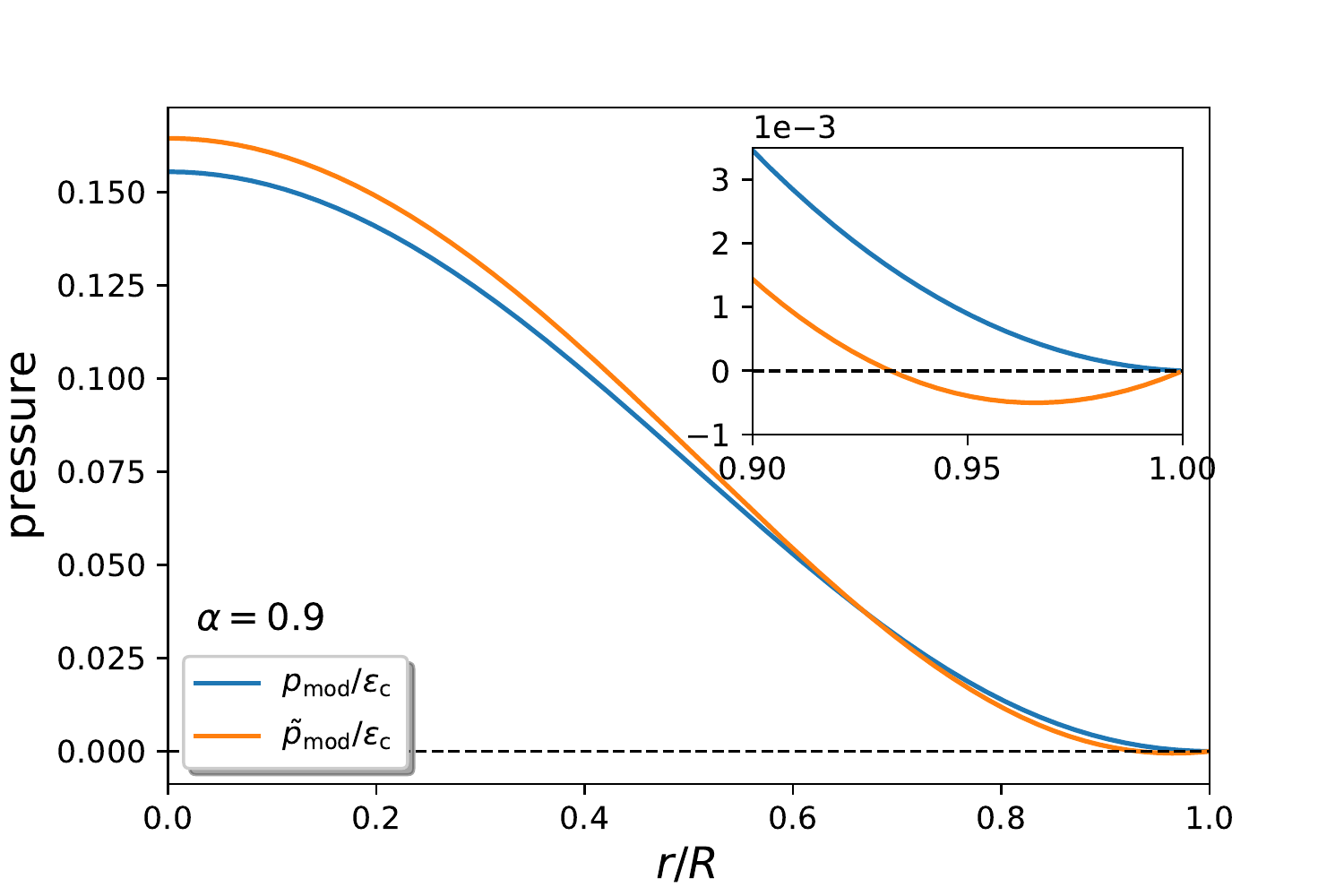}
    \caption{Profiles of the pressure $\tilde{p}_\text{mod}$ [Eq.~\eqref{pOrig}] and the corrected pressure $p_\text{mod}$ [Eq.~\eqref{pMod}], as a function of $r/R$, for the MTVII solution. We consider a NS with $M=1.4\,M_{\odot}$, $R=11.4\,\text{km}$, and $\alpha \in [0, 1)$. The pressure is measured in units of the central energy density $\epsilon_\mathrm{c}$. Notice the region of negative pressure near the surface predicted by the original $\tilde{p}_\mathrm{mod}$.}
\label{fig:1}
\end{figure*}

Let us discuss now some inconsistencies we found with the expression for the corrected pressure $p_{\imp}$ [Eq.~\eqref{pMod}]. Following~\cite{Jiang2019}, we consider a NS with $M=1.4~M_{\odot}$ and $R=11.4~\text{km}$. In Fig.~\ref{fig:1} we show the profiles of $\tilde{p}_{\imp}$ and $p_{\imp}$, as a function of $r/R$, for some values of $\alpha \in [0, 1)$. We observe that $\tilde{p}_{\imp}$ becomes negative near the surface of the configuration, thus confirming the findings reported by~\cite{Jiang2019}. The extreme case is given by $\alpha = 0$, where the negative pressure region comprises almost 20\% of the star. However, as $\alpha$ increases the negative pressure region decreases, and it vanishes for the original TVII solution ($\alpha=1$) where $\tilde{p}_{\imp}=p_{\imp}$. In contrast, we observe that the corrected pressure $p_{\text{mod}}$ is positive in the whole interior of the star in this range of $\alpha$.

\begin{figure*}
    \includegraphics[width=.44\linewidth]{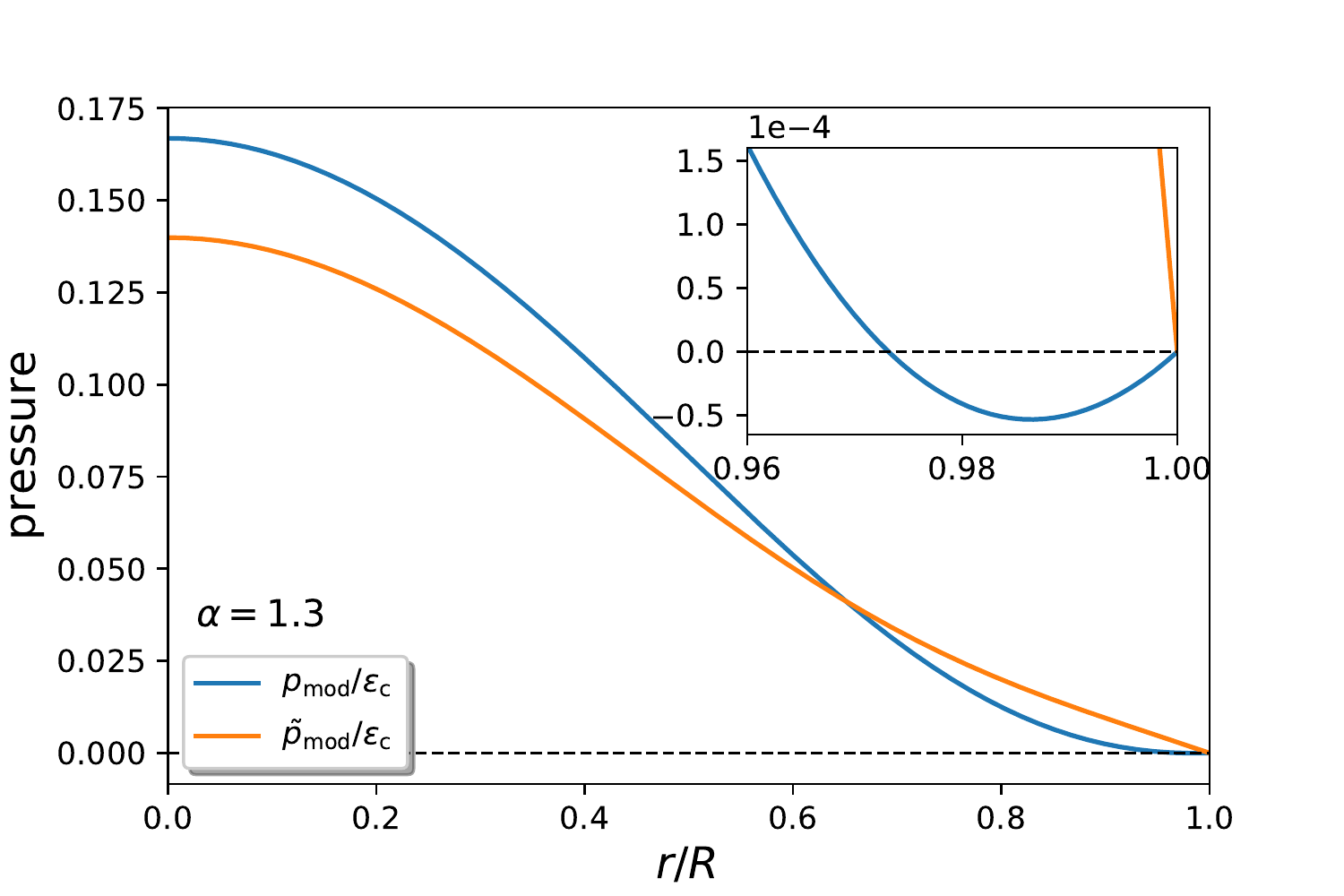}\hfill
    \includegraphics[width=.44\linewidth]{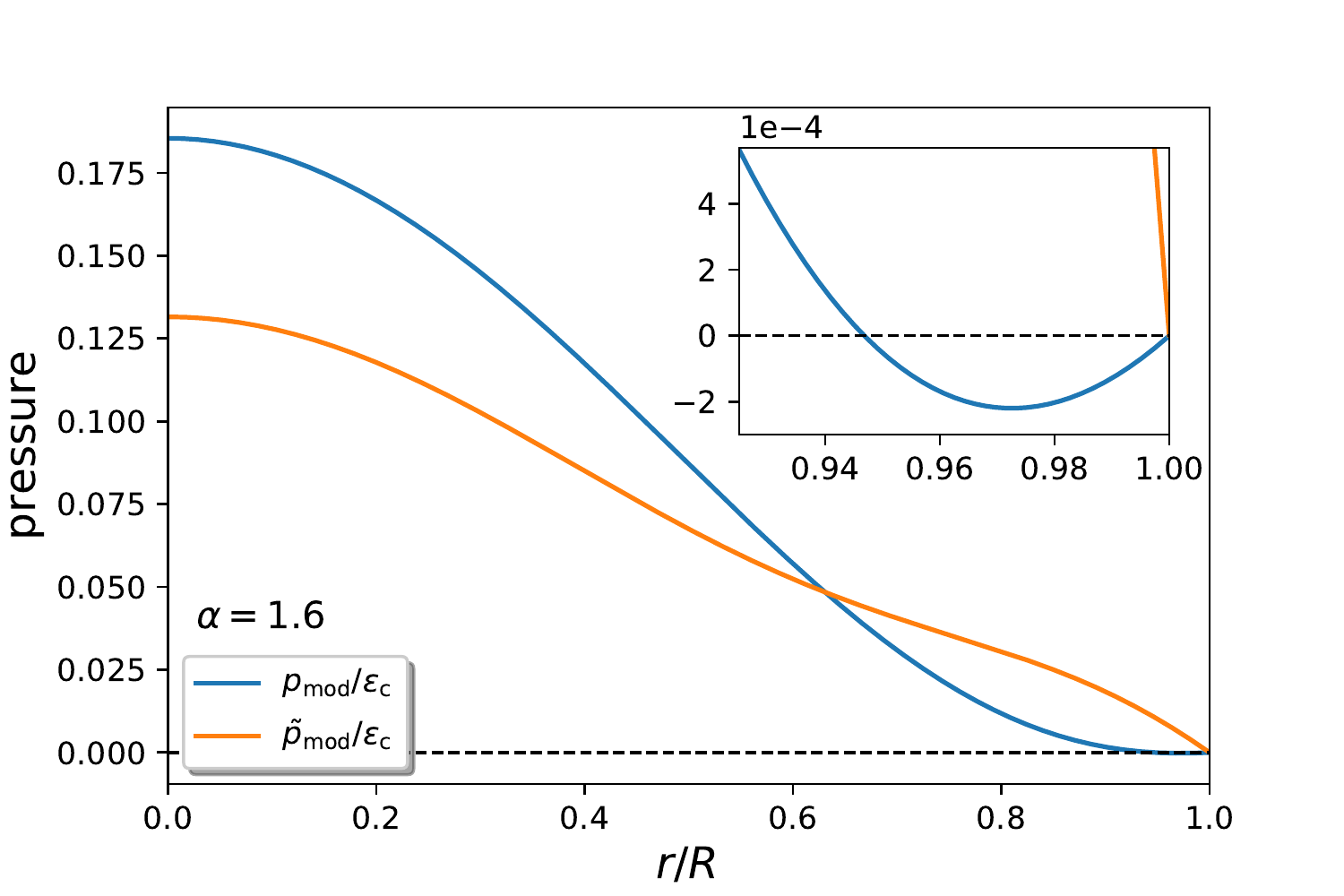}
    \includegraphics[width=.44\linewidth]{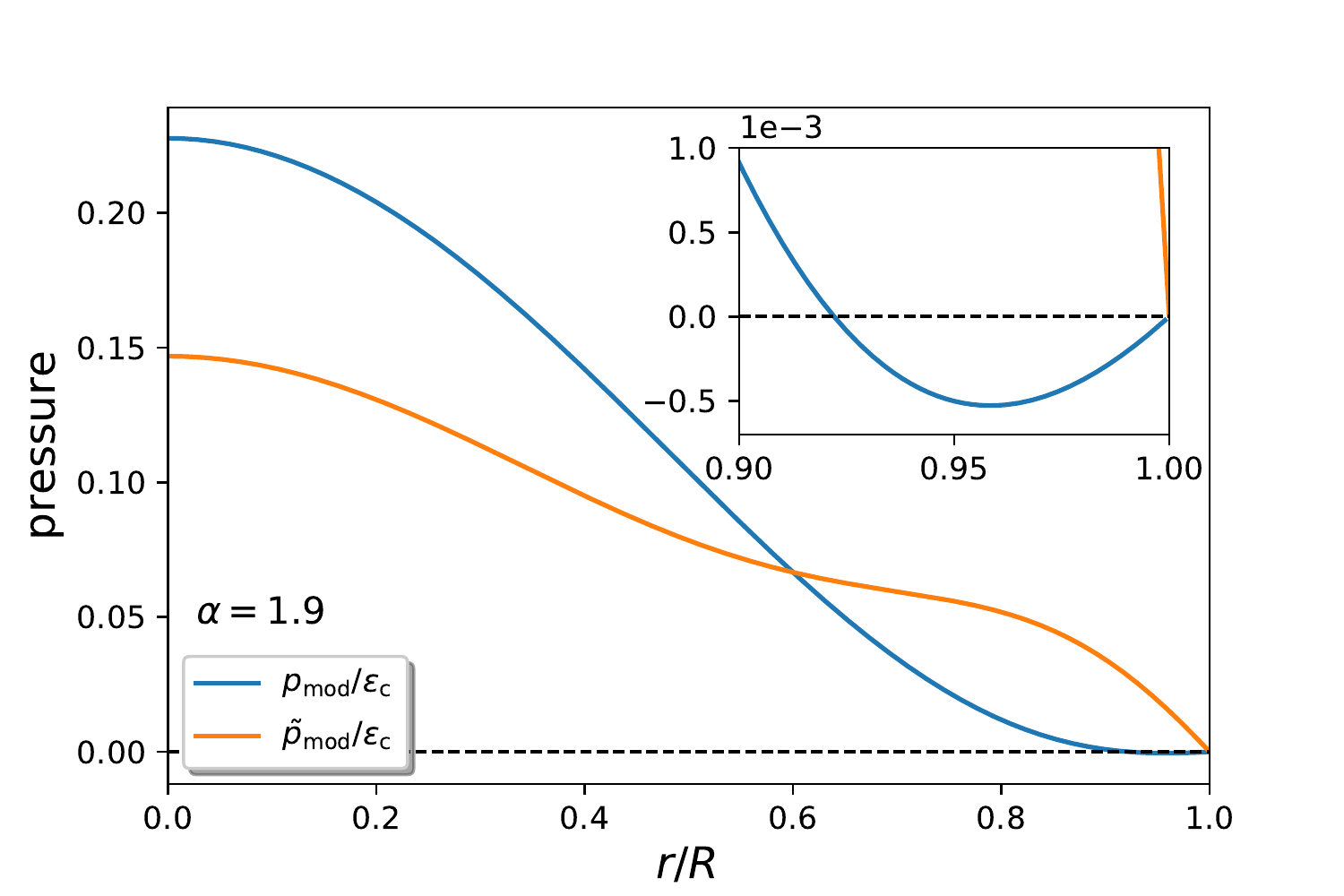}\hfill
    \includegraphics[width=.44\linewidth]{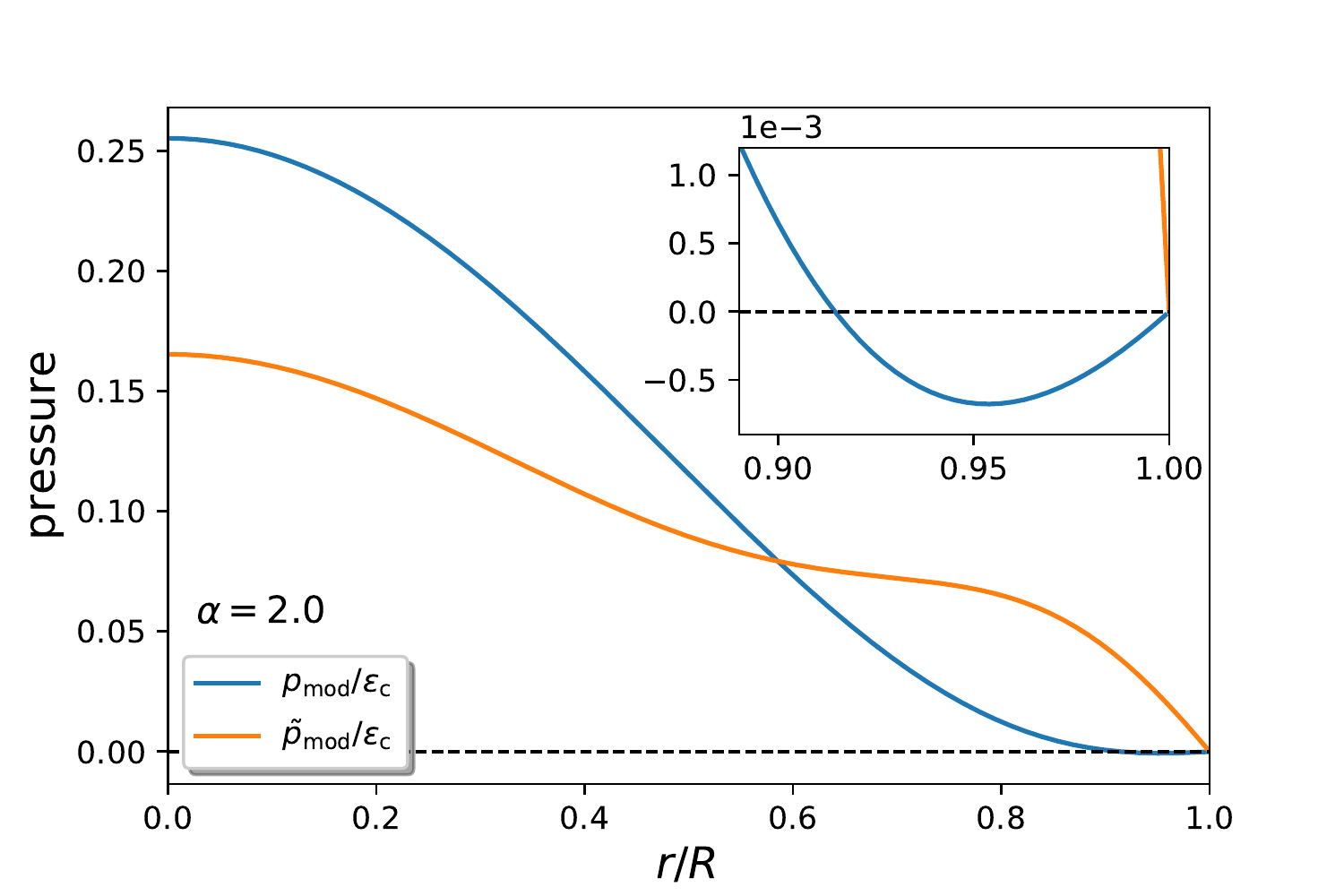}
    \caption{Same profiles as Fig.~\ref{fig:1}, but now $\alpha \in (1, 2]$. Note that the corrected pressure $p_\mathrm{mod}$ shows a region of negative pressure near the surface. Observe that for $\alpha=2$ the negative pressure region comprises almost $10\%$ of the star.}
\label{fig:2} 
\end{figure*}

As we show in Fig.~\ref{fig:2}, when we consider $\alpha$ in the range $\alpha \in (1, 2]$, we found that the arguments drawn in~\cite{Jiang2019} regarding the pressure are flawed. First of all, we observe that $\tilde{p}_\mathrm{mod}$ is positive throughout the star, however, note its peculiar behavior for the cases $\alpha=\{1.9, 2\}$, which shows a plateau at intermediate values of $r$, where the pressure is nearly constant, and then rapidly decreases. From the relativistic hydrostatic equilibrium condition (Tolman-Oppenheimer-Volkoff equation), 

\begin{equation}
\frac{dp}{dr}=-\frac{m(r)\epsilon(r)}{r^2}\left(1+\frac{p(r)}{\epsilon(r)}\right)\left(1+\frac{4\pi r^3 p(r)}{m(r)}\right)e^{\lambda(r)}, 
\end{equation}

\noindent together with the mass continuity relation [Eq.~\eqref{einstein3}] and the profile of the energy density [Eq.~\eqref{rhoMod}], which does not show any plateau, it is not expected to see this behavior in the pressure. On the other hand, the corrected expression for the pressure $p_\mathrm{mod}$, which was introduced by~\cite{Jiang2019} to alleviate the negative pressure behavior of $\tilde{p}_\mathrm{mod}$ near the surface, turns out to be negative near the surface. We have included an inset enlargement of the negative pressure region. We observe that as $\alpha$ increases, the negative pressure region also increases, reaching almost $10\%$ of the star when $\alpha=2$.

\begin{figure}
    \includegraphics[width=\columnwidth]{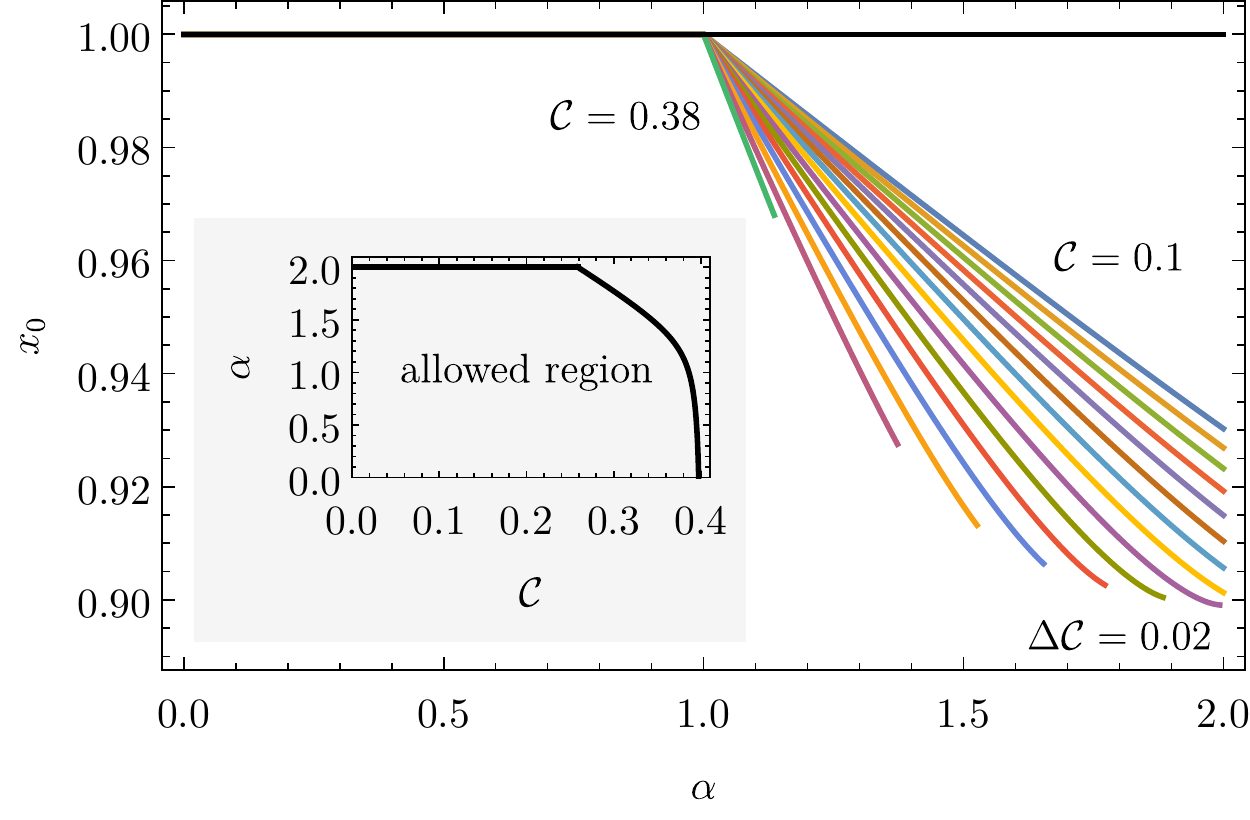}
    \caption{\label{roots} Locations $x_{0}$ where $p_{\imp}(x_0)=0$, as a function of the parameter $\alpha$. We consider the compactness in the range $\com\in[0.1,0.38]$, with a step size 0.02 (different values of $\com$ are depicted in different colors). Note that $p_{\imp}=0$ for $x = 1$ for each case. However, for a given compactness, when $\alpha>1$ there exists also a second root $x_{0}<1$. Inset: we show the constraints on the values for $(\com,\alpha)$ as given by the condition of finiteness of the central pressure.}
\label{fig:3}
\end{figure}

We determined the roots $x_0$ corresponding to the values of $x$ where the corrected pressure vanishes, i.e., $p_{\text{mod}}(x_0)=0$, for $\alpha\in[0,2]$ and $\com\in[0.1,0.38]$. We present our results in Fig.~\ref{roots}. Note that $x_0=1$ in the regime $\alpha\in[0,1]$ which is consistent with the expected radius of the star; however, when $\alpha>1$ there appears a second root, $x_{0}<1$, indicating the origin of the negative pressure region. Note that for values of $\alpha \in (1, 1.6)$, with $\com<0.2$, the region of negative pressure is relatively small (roughly less than 5\% of the star); however, for bigger values of $\alpha$ it increases, taking its maximum value, corresponding to almost 10\% of the star, when $\alpha > 1.6$ with $\com>0.3$.

As we will show in Sec.~\ref{sec:5}, this peculiar behavior of the pressure $p_{\imp}$ from the MTVII model, led us to find that the MTVII solution predicts negative tidal deformability, for a certain combination of the parameters $(\mathcal{C},\alpha)$, which is not consistent with previous results for realistic NSs \cite{Hinderer2008, Hinderer:2009ca, Damour2009, Postnikov2010}. To alleviate these issues, in the following section, we present an exact solution to Einstein's equations, for the modified density profile $\epsilon_{\text{mod}}$.

\section{Exact Modified TVII solution}\label{sec:4}

In this section, we present an exact solution to Einstein's equations for the quartic energy density model $\epsilon_\imp$, or exact modified Tolman VII solution. As it was pointed out in \cite{Jiang2019}, it is not possible to find a complete analytical solution for this density profile. Therefore, we solved numerically Einstein's equations [Eqs.~\eqref{einstein1}--\eqref{einstein3}] in order to determine the metric function $g_{tt}$ and the radial pressure $p$ (note that the metric function $g_{rr}$ already satisfies exactly the Einstein equations). Even though ours is a numerical solution, it is exact in the sense that we are not assuming any approximation for the metric component $g_{tt}$ or the pressure profile as it was done by \cite{Jiang2019}. We only assume the quartic energy density profile $\epsilon_{\text{mod}}$.

\begin{figure}
    \includegraphics[width=\columnwidth]{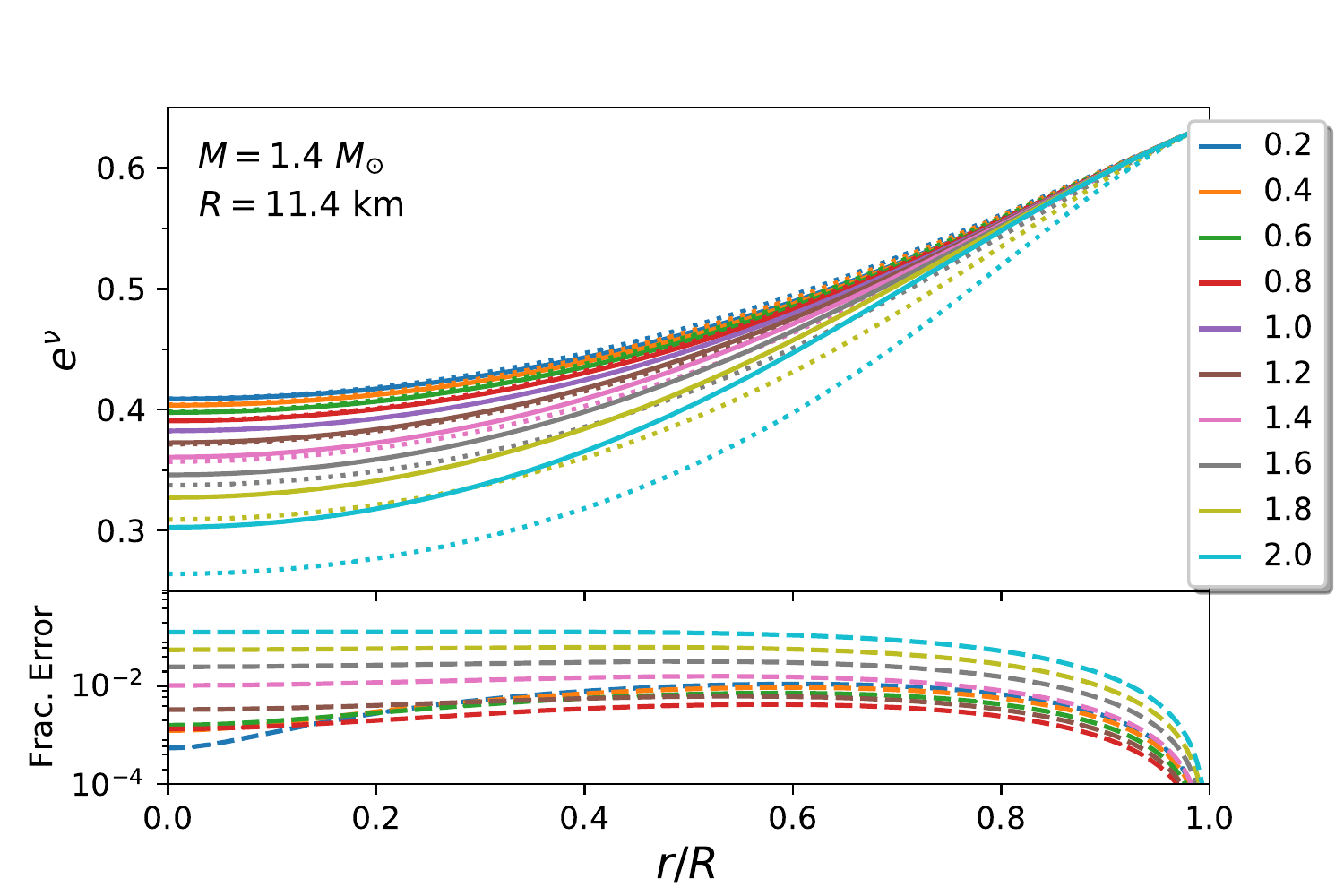}
    \caption{Time metric component $g_{tt}$, as a function of the radial coordinate $r$ (in units of $R$), for the EMTVII (solid lines) and the MTVII (dotted lines) solutions. We consider $\alpha\in[0,2]$. The bottom shows the fractional error.}
\label{fig:4}
\end{figure}

In Fig.~\ref{fig:4} we show the profiles for the time metric component $g_{tt}\equiv e^{\nu}$, as a function of the radial coordinate $r/R$, for the EMTVII (solid line) and MTVII (dotted line) solutions, for some representative values of $\alpha$ in the full range $\alpha\in[0,2]$. We indicate with the same color the configurations with the same value of $\alpha$. The lower panel shows the corresponding fractional error. We observe that as $\alpha$ increases, the $g_{tt}$ component decreases. Note that the various curves tend to the same value as they approach the surface, where they match smoothly with the exterior Schwarzschild spacetime. We observe that the differences are the biggest in the core and the central region of the star and decrease as $r\to R$. For configurations with $\alpha\geq 1.4$, the fractional error for the central value of $g_{tt}$ is above $1\%$, going up to $12\%$ for $\alpha=2$. 

\begin{figure}
    \includegraphics[width=\columnwidth]{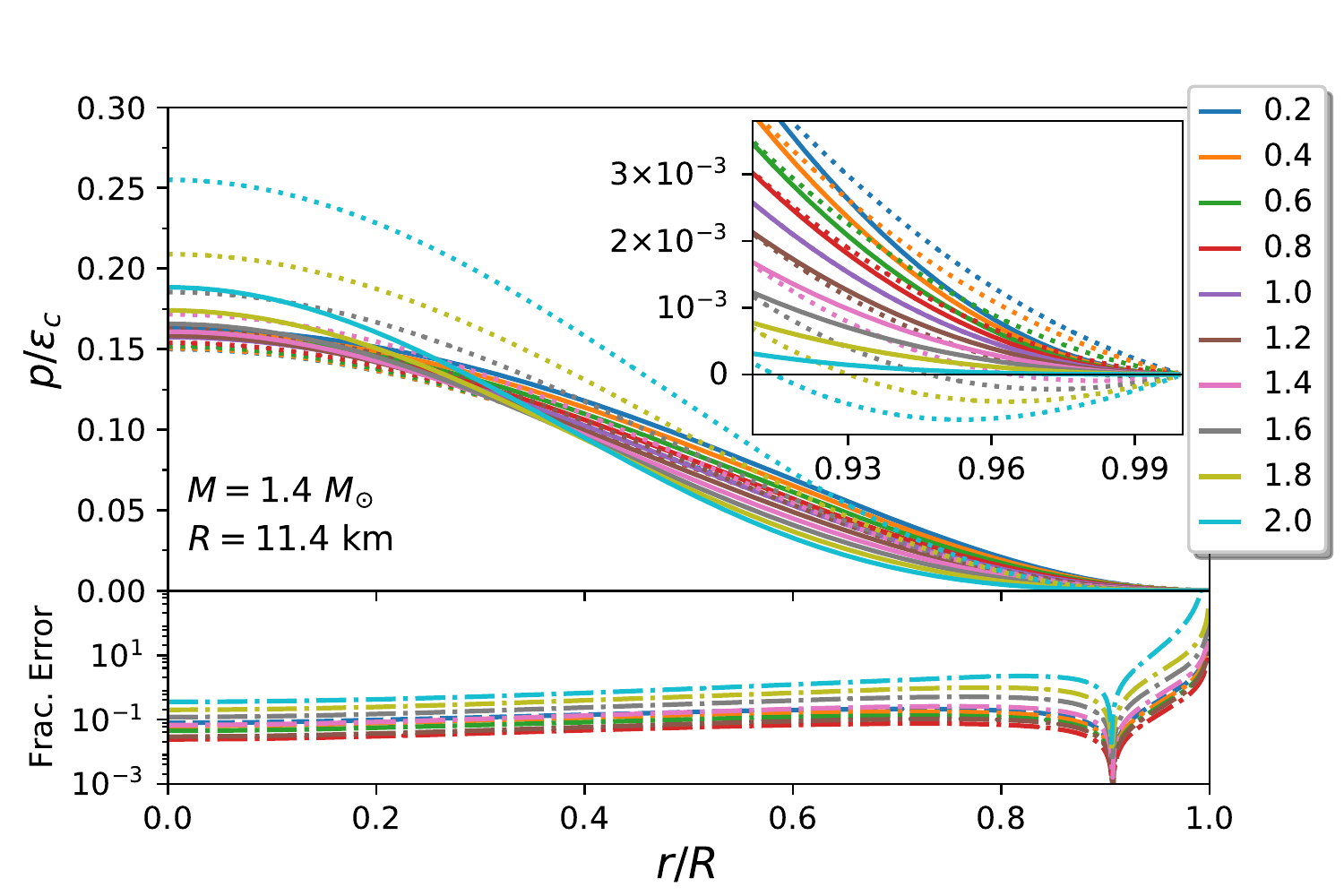}
    \caption{Profiles of the radial pressure (in units of $\epsilon_{\text{c}}$) for the EMTVII (solid lines) and MTVII (dotted lines) solutions, for $\alpha\in[0,2]$. Note that for $\alpha>1$, the MTVII pressure $p_{\imp}$ shows a region of negative pressure near the surface. Meanwhile, the EMTVII pressure is positive throughout the star. The bottom shows the fractional error.}
\label{fig:5}
\end{figure}

In Fig.~\ref{fig:5} we present the profiles for the radial pressure of the EMTVII (solid lines) and MTVII (dotted lines) solutions. The pressure is being measured in units of the central energy density $\epsilon_{\text{c}}$. The lower panel shows the corresponding fractional errors.  
First of all, we observe that the pressure for the EMTVII solution is positive throughout the star, for all the values of $\alpha$, and decreases monotonically with $r$. As it was discussed in the previous section, note that for the MTVII solution the pressure $p_{\imp}$ becomes negative near the surface for values of $\alpha>1$. In consequence, the fractional error increases dramatically in this region. For values of $\alpha<1$, the fractional errors in the central pressure are below $10\%$; meanwhile, in the regime where $\alpha>1$, the fractional errors are ranging from $3\%$ for $\alpha=1.2$ up to near $35\%$ for $\alpha=2$.     

\begin{figure}
    \includegraphics[width=\columnwidth]{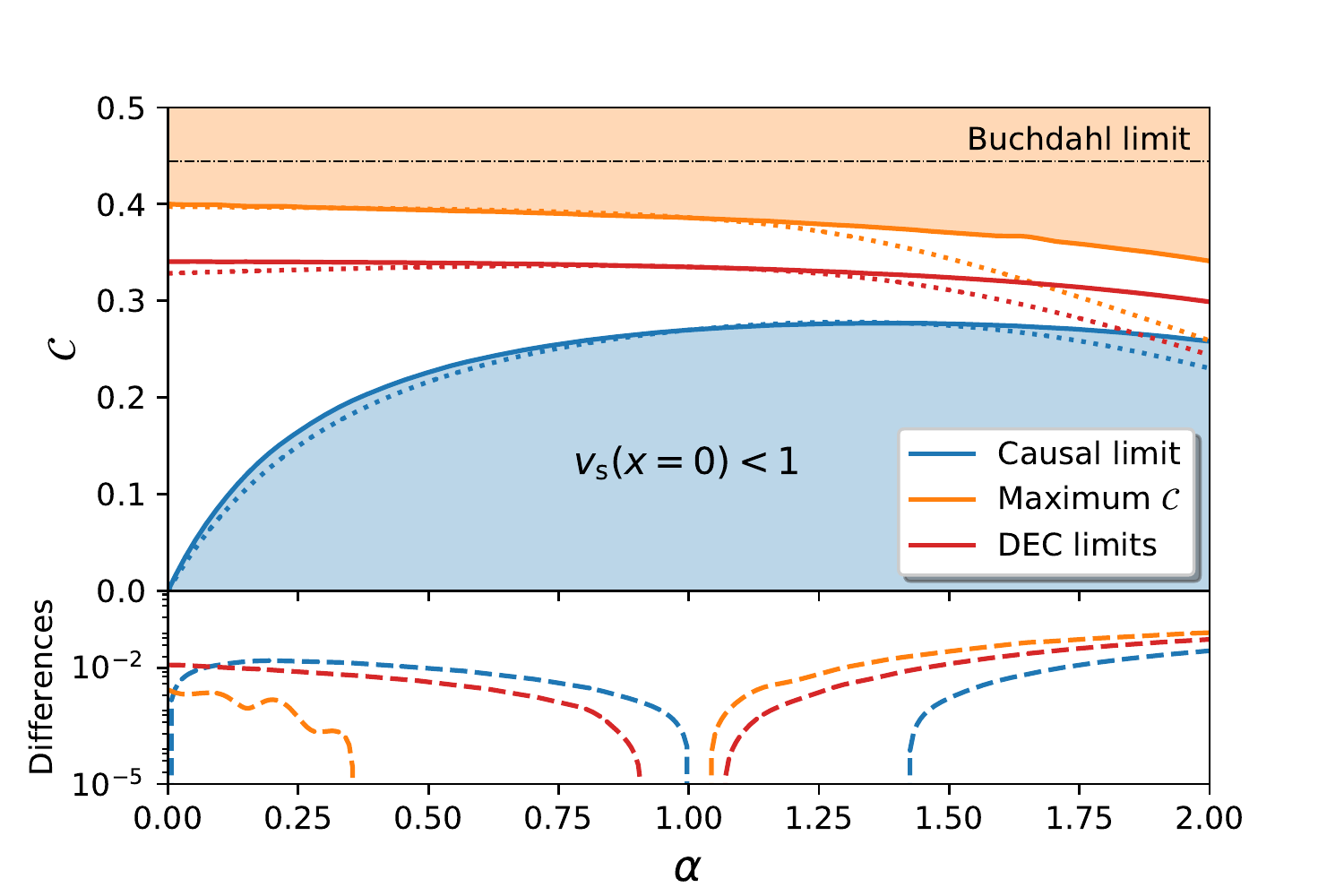}
    \caption{Constraints on the compactness $\mathcal{C}$, for $\alpha\in[0,2]$, predicted by the EMTVII (solid lines) and MTVII (dotted lines) solutions. The blue lines indicate the causal limit determined by the condition $v_{\text{sound}}(x=0)\leq1$. The orange lines indicate the maximum compactness where the central pressure diverges. The red lines correspond to the limits imposed by the DEC. The bottom shows the fractional error.}
    \label{fig:6}
\end{figure}

In Fig.~\ref{fig:6} we show the causality, finite central pressure, and DEC domains in the parameter space $(\mathcal{C},\alpha)$, for the whole allowed range $\alpha\in[0,2]$, for the EMTVII (solid lines) and the MTVII (dotted lines) solutions. The blue lines delimit the causality domain as given by the condition that the central speed of sound must be subluminal, $(\partial p/\partial \epsilon)_{r=0}<1$. The orange curves indicate the maximum compactness $\com_{\text{max}}$ such that the central pressure is finite. The red curves correspond to the limits established by the DEC. We observe how the parameter $\alpha$ restricts the solution, in the sense that as $\alpha$ increases the allowed $\com_{\text{max}}$ decreases. Observe that for $\alpha<1$, both solutions predict almost the same maximum $\com$, with fractional differences below $1\%$. However, for $\alpha>1$, the limits on $\com_{\text{max}}$ predicted by the EMTVII solution are less restrictive than those obtained with MTVII. Fractional errors go from near $10\%$ for $\alpha=1.6$ up to $24\%$ for $\alpha=2$.

\begin{figure}
    \includegraphics[width=\columnwidth]{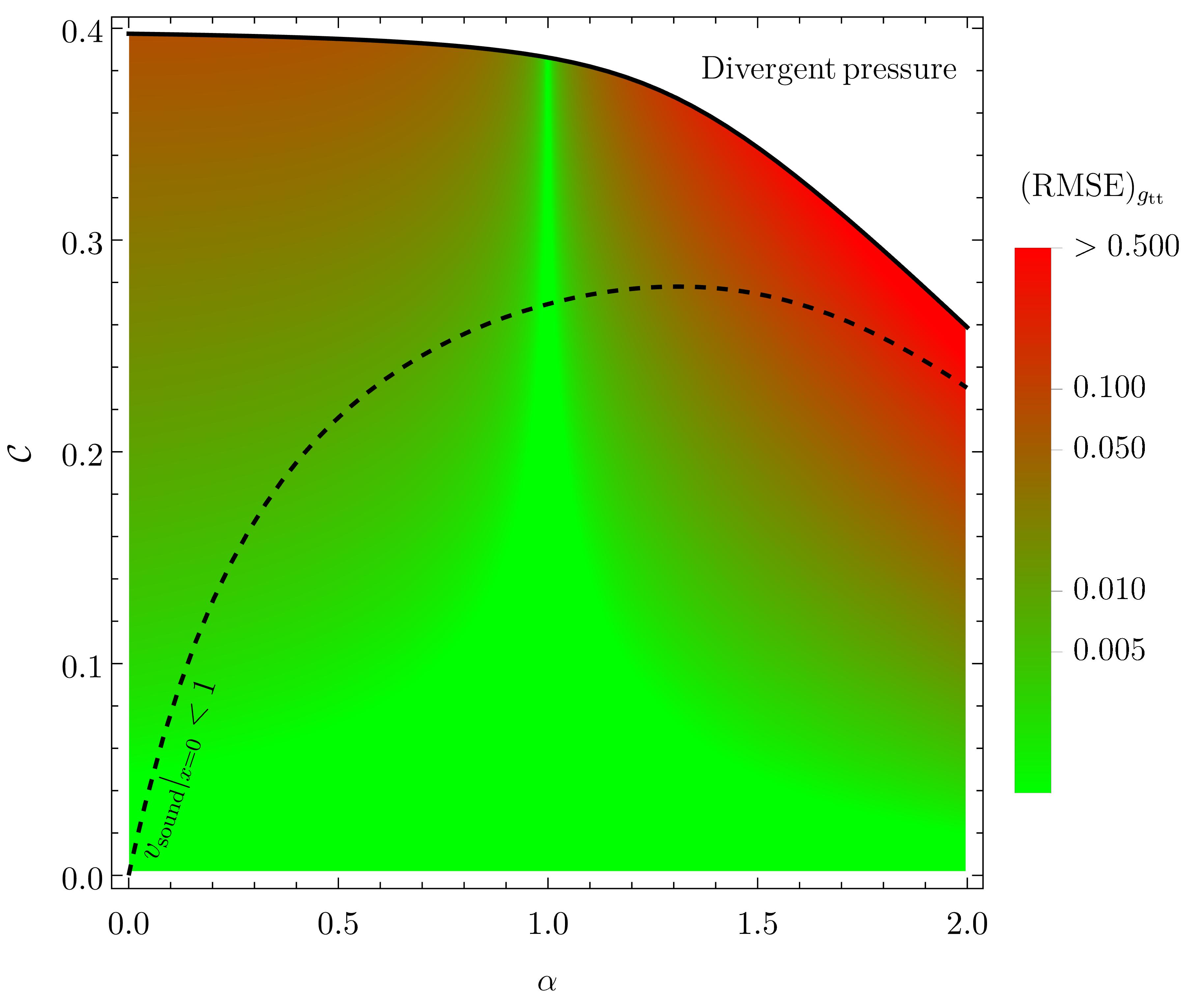}
    \caption{The relative RMSE [see Eq.~\eqref{rmse}] of the time metric component $g_{tt}$, for the EMTVII and MTVII solutions in the whole parameter space $(\alpha,\com)$.}
\label{fig:7}
\end{figure}

\begin{figure}
    \includegraphics[width=\columnwidth]{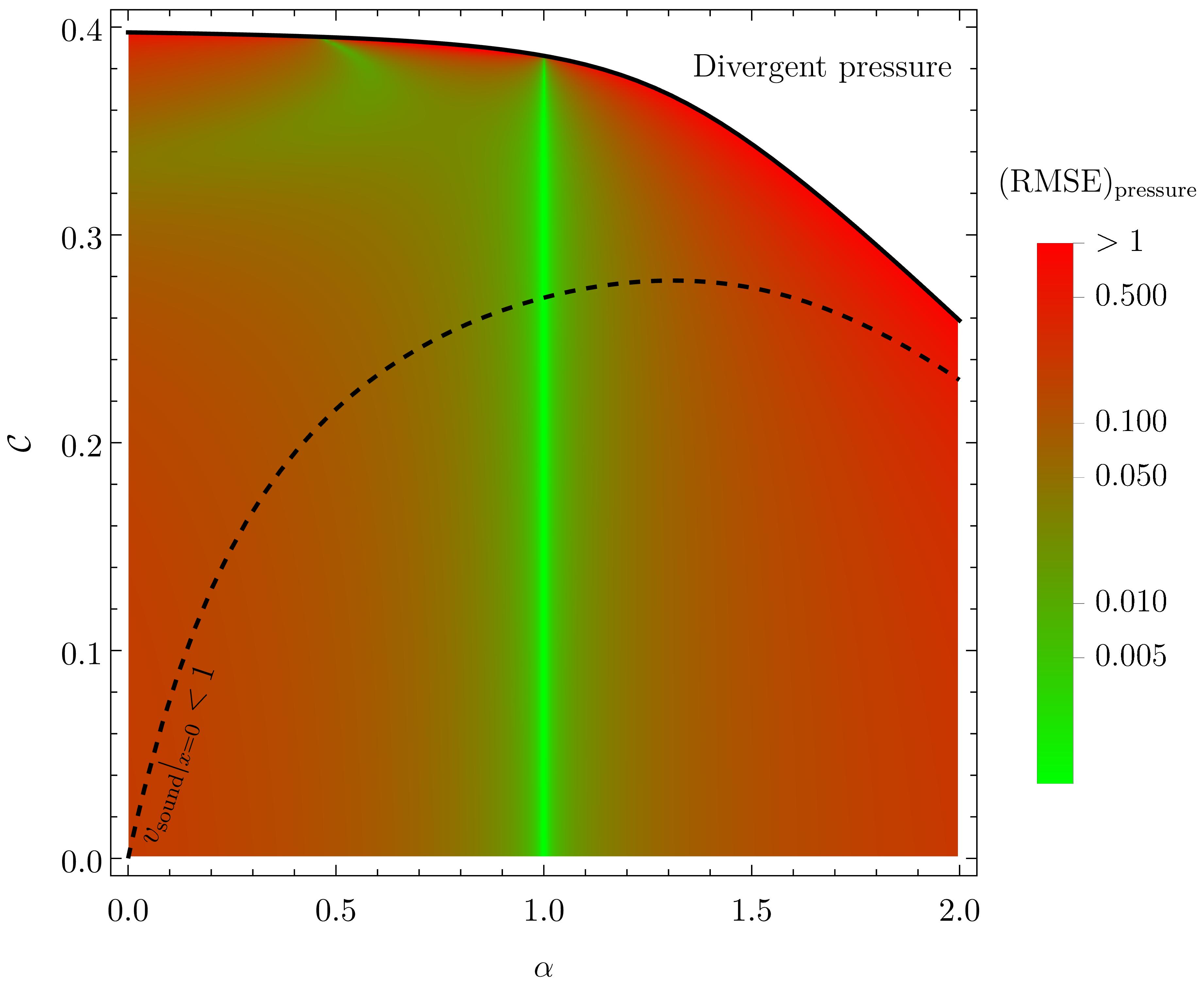}
    \caption{The relative RMSE [see Eq.~\eqref{rmse}] of the radial pressure profile, for the EMTVII and MTVII solutions in the whole parameter space $(\alpha,\com)$.}
\label{fig:8}
\end{figure}

The results presented so far are specific to one configuration. Therefore, in Figs.~\ref{fig:7} and \ref{fig:8}, we show the differences, in the whole parameter space $(\com,\alpha)$, for the metric time component $g_{tt}$ between the numerical EMTVII solution and the analytical MTVII model. Following \cite{Jiang2019} we employ the relative root-mean-square error (RMSE), which gives a measure of the error between the analytical model and the numerical results and is defined as

\begin{equation}\label{rmse}
(\text{RMSE}) = \sqrt{\frac{\int_0^R \left[y^{\text{(EMTVII)}} - y^{\text{(MTVII)}}\right]^2\,dr}{\int_0^R [y^{\text{(EMTVII)}}]^2\,dr}}\,,
\end{equation}

\noindent where $y=(g_{tt}, p)$ and the integration is carried out throughout the star. From the results depicted in Fig.~\ref{fig:7}, we observe that acceptable differences (up to a few percent) are obtained for values of $\alpha$ close to $1$, for any compactness. Note that for $\alpha=1$, i.e., the original TVII solution, there appears a peak indicating that both models provide the same predictions. In the low compactness regime, $\com <0.15$, the analytical MTVII model seems to be a good approximation for the whole span of $\alpha$ values. However, for certain combinations of the parameters $(\com,\alpha)$, the analytical MTVII $g_{tt}$ profile starts to deviate from the EMTVII solution. For instance, for the range considered in~\cite{Jiang2020}, namely, $\alpha\in[0.4,1.4]$ with $\com\in[0.05,0.35]$, we observe that configurations with $\alpha<0.5$ and $\com>0.3$ are already in the orange region where the RMSE is above 0.05. Moreover, configurations with $\alpha=1.4$ and $\com>0.3$ enter into the red region where the RMSE is close to 0.5. In any case, we are restricted by the causality condition, i.e., the central speed of sound is less than 1 (see Fig.~\ref{fig:6}). The speed of sound in the interior of the star decreases monotonically starting from its maximum value at the center and approaching zero at the surface. We have included the causality (black dotted line) and finite central pressure (black solid line) limits, determined by the MTVII model, given that these are more restrictive than those predicted by the EMTVII solution; as a consequence, it is not possible to determine differences for configurations beyond the limits set by the MTVII model. 

Similar to Fig.~\ref{fig:7}, in Fig.~\ref{fig:8} we show the RMSE for the pressure profiles, for the EMTVII and MTVII solutions. We observe that the MTVII pressure approximation deviates considerably, for a wide range of values $(\com,\alpha)$, from the exact solution to Einstein equation. For values of $\alpha\to1$, both models are in good agreement as expected.  

As a further application of our new EMTVII solution, in the following sections, we study the tidal deformability and the tidal Love number $k_2$ and compare it with the values predicted by the MTVII model. 

\section{Tidal deformability and Love numbers}\label{sec:5}

The tidal Love number $k_2$, or alternatively the tidal deformability $\Lambda$, provides a connection between the external tidal quadrupole moments $\varepsilon_{ij}\equiv -\partial_{ij}U^{\text{ext}}$, given by second derivatives of the external potential $U^{\text{ext}}$ generated by the distant objects producing the tidal effects and the induced mass quadrupole moments $Q_{ij}$. This relation is given by~\cite{Hinderer2008,Damour2009}
\begin{equation}\label{tidal}
  Q_{ij} = - \frac{2k_2 R^5}{3} \varepsilon_{ij} \equiv - \Lambda \varepsilon_{ij}\, .
\end{equation}
It is worthwhile to recall that Eq.~\eqref{tidal} neglects nonlinear and time-dependent terms, plus additional post-Newtonian corrections~\cite{Poisson:2021yau}. The tidal Love number depends on the EOS of the star, and it is also strongly sensitive to the compactness. It is conventional to introduce the dimensionless tidal deformability
\begin{equation}\label{Lambda}
  \bar{\Lambda} = \Lambda / M^5 = 2k_2/(3\mathcal{C}^5)\, ,
\end{equation}
which is frequently used in the context of the I-Love-Q relations for NSs~\cite{Yagi:2013awa}. Following the notation and formulation of Ref.~\cite{Damour2009}, the even-parity metric perturbation $H = H_0 = H_2$ satisfies a single second-order differential equation given by
\begin{equation}\label{perturb}
  \frac{\mathrm{d}^2 H}{\mathrm{d}r^2} + C_1(r)\frac{\mathrm{d}H}{\mathrm{d}r} + C_0(r) H = 0\, ,
\end{equation}
where the coefficients $C_0$ and $C_1$ are
\begin{multline}\label{coeftid0}
  C_0(r) = \ee^\lambda \left[-\frac{l(l+1)}{r^2}+4\pi(\epsilon+p)\frac{\mathrm{d}\epsilon}{\mathrm{d}p}+4\pi(5\epsilon+9p) \right] - \\ \left(\frac{\mathrm{d}\nu}{\mathrm{d}r}\right)^2\, ,
\end{multline}
\begin{equation}\label{coeftid1}
  C_1(r) = \frac{2}{r} + \ee^\lambda \left[-\frac{2m}{r^2}+4\pi(p-\epsilon)\right]\, ,
\end{equation}
with $l$ denoting the multipole order. Using the logarithmic derivative $h(r) \equiv (r/H)\mathrm{d}H/\mathrm{d}r$, Eq.~\eqref{perturb} can be  rewritten as a Riccati-type equation
\begin{equation}\label{ricatti}
  r\frac{\mathrm{d}h}{\mathrm{d}r} + h(h - 1) + r C_1 h + r^2 C_0 = 0\, ,
\end{equation}
with the regular solution near the origin
\begin{equation}\label{hr0}
  h(r) = l.
\end{equation}
The tidal Love number $k_2$ can be obtained from the following expression
\begin{multline}\label{k2}
  k_2(\mathcal{C}, h_R) = \frac{8}{5} (1 - 2\mathcal{C})^2 \mathcal{C}^5 \left[2\mathcal{C}(h_R - 1) - h_R + 2\right]\times \\ \left\{2\mathcal{C}[4(h_R + 1)\mathcal{C}^4+(6h_R - 4)\mathcal{C}^3 + (26 - 22h_R)\mathcal{C}^2 + \right.\\ \left. 3(5h_R - 8)\mathcal{C} - 3h_R + 6] + 3(1 - 2\mathcal{C})^2 \right. \times \\ \left. [2\mathcal{C}(h_R - 1) - h_R + 2]\log(1 - 2\mathcal{C})\right\}^{-1}\, ,
\end{multline}
where $h_R$ is the value of $h$ at the surface $r = R$.

\subsection{Numerical results}

\begin{figure}
    \includegraphics[width=\columnwidth]{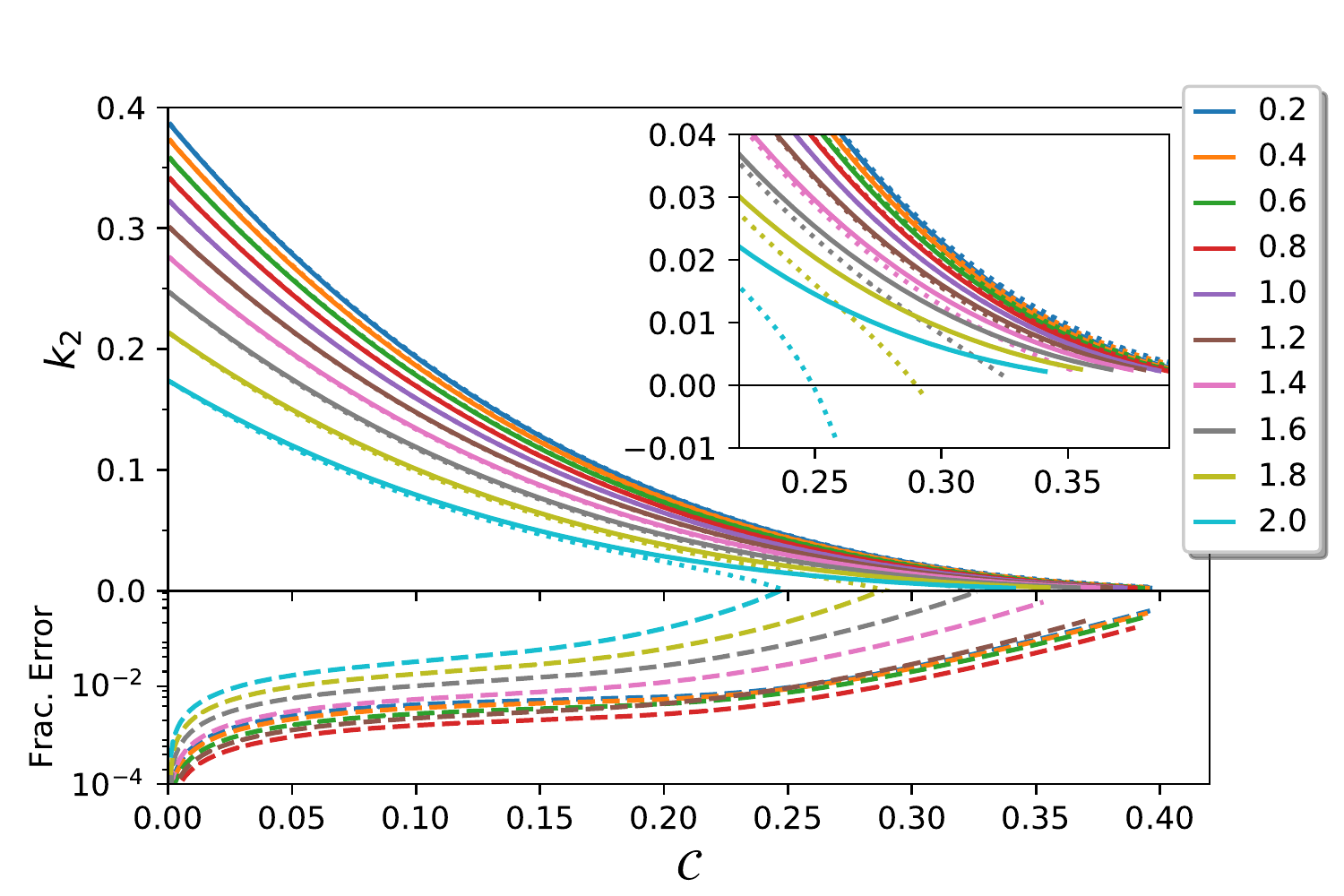}
    \caption{Tidal Love number $k_2$, as a function of the compactness, for the EMTVII (solid lines) and MTVII (dotted lines) solutions, for $\alpha\in[0,2]$. The bottom shows the fractional error. Observe that the MTVII solution predicts negative tidal deformability for values of $\alpha>1.6$ and $\com>0.25$.}
\label{fig:9}
\end{figure}

In Fig.~\ref{fig:9} we present the results of the tidal Love number $k_2$ for the EMTVII (solid lines) and MTVII (dotted lines) solutions, for various values of $\alpha$ in the range $\alpha\in[0,2]$. In the lower panel, we show the corresponding fractional errors. First of all, for the original TVII solution ($\alpha=1$), our results are in very good agreement with those reported in~\cite{Postnikov2010}. We observe that for the EMTVII solution, $k_2$ is continuous and decreases monotonically with $\com$. Moreover, as the compactness increases, the different curves approach each other. Furthermore, as the configurations approach the black hole (BH) compactness limit, i.e., $\com=1/2$, $k_2\to 0$ corresponding to the BH value \footnote{Strictly speaking, the BH limit cannot be taken from these particular EoSs. However, it has been shown that ultracompact stars with uniform density, or Schwarzschild stars, can evade the Buchdahl bound and approach the BH compactness limit \cite{Chirenti:2020bas}. Moreover, in this limit these configurations show zero tidal deformability, making them a good candidate as a BH mimicker.}. This behavior is in good agreement with the one found for the EOS of realistic NSs reported by \cite{Damour2009, Hinderer:2009ca}. However, in the low compactness regime, for realistic EOS, $k_2\to 0$ as $\com\to 0$ (see Fig. 1 in \cite{Hinderer:2009ca}); meanwhile the EMTVII solution predicts that the tidal Love number ranges from $k_2\sim0.4$ for $\alpha\to 0$ down to $k_2\sim 0.16$ for $\alpha=2$. 

For the MTVII solution, we observe a similar behavior of $k_2$ as for the EMTVII model, at least up to $\alpha=1.2$. However, for values of $\alpha>1.2$, there appears a sharp contrast between both models. For instance, as the compactness increases, note that the different curves for $k_2$ predicted by the MTVII solution \emph{deviate} from each other. Moreover, for configurations with $\alpha$ above $1.6$ and compactness above $\sim 0.25$, the Love number drops rapidly to zero and becomes negative. This behavior is not consistent with what is expected for the realistic NSs~\cite{Damour2009, Hinderer:2009ca}, which are characterized by positive tidal deformability \footnote{Certain exotic compact objects (ECOs), like thin-shell gravastars, wormholes, anisotropic stars, among others, have been shown to have negative tidal deformability \cite{Cardoso2017}. However, an NS is not considered an ECO in this context.}. 

For the NS we have been considering so far, with $M=1.4~M\odot$ and $R=11.4~\text{km}$, from Fig. \ref{tidal} we observe that fractional errors go from less than $1\%$ for $\alpha\to 0$, up to $\sim 10\%$ for $\alpha=2$. Let us consider the regime of $\alpha$ restricted by \cite{Jiang2020}, i.e., $\alpha\in[0.4,1.4]$ with $\com\in[0.05,0.35]$. For instance, for a configuration with $\com\sim 0.32$ and $\alpha=0.6$, which matches with a WFF1 EOS (see Fig. 2 in \cite{Jiang2020}), the fractional error is around $3\%$. For $\alpha=1.4$ with $\com\sim 0.25$ (Shen EoS) the fractional error is close to $2.8\%$. Thus, in principle, in the restricted regime considered by~\cite{Jiang2020} the values of the tidal Love number for the MTVII model are relatively close to those predicted by the EMTVII solution. Nevertheless, given the ignorance in the realistic EOS for a NS, there is no reason to discard a configuration that lies in the regime above $\alpha=1.6$ with $\com>0.25$. For such a case, the differences grow significatively, without mentioning the peculiar negative values for $k_2$ predicted by the MTVII model. Therefore it is worthwhile to consider the whole range $\alpha\in[0,2]$, using the new EMTVII solution, as we have already discussed.    

\subsection{Observational implications}

In this section, we translate our results to $\bar{\Lambda}$ to include current constraints on the tidal deformability obtained from the binary NS merger event GW170817~\cite{De:2018uhw, Flanagan:2007ix}. It is convenient to introduce the binary tidal deformability  $\tilde{\Lambda}$ defined as 

\begin{equation}\label{binary_tidal}
\tilde{\Lambda}=\frac{16}{13}\frac{(1+12q)\bar{\Lambda}_1 + (12+q)q^{4}\bar{\Lambda}_2}{(1 + q)^5},
\end{equation}   

\noindent which gives a mass-weighted average of the dimensionless tidal deformabilities $\bar{\Lambda}_{1}$ and $\bar{\Lambda}_{2}$, where $q=M_2/M_1$ denotes the binary mass ratio. For an equal mass binary, $q=1$, we have $\tilde{\Lambda}=\bar{\Lambda}_1=\bar{\Lambda}_2$. The event GW170817 set the constraints $50 \lesssim\tilde{\Lambda} \lesssim 800$ and also  $0.7<q<1$, which implies $2.73 M_{\odot}<M_1 + M_2 < 3.05 M_{\odot}$, at the $90\%$ confidence level~\cite{LIGOScientific:2018hze}. 

\begin{figure}
    \includegraphics[width=\columnwidth]{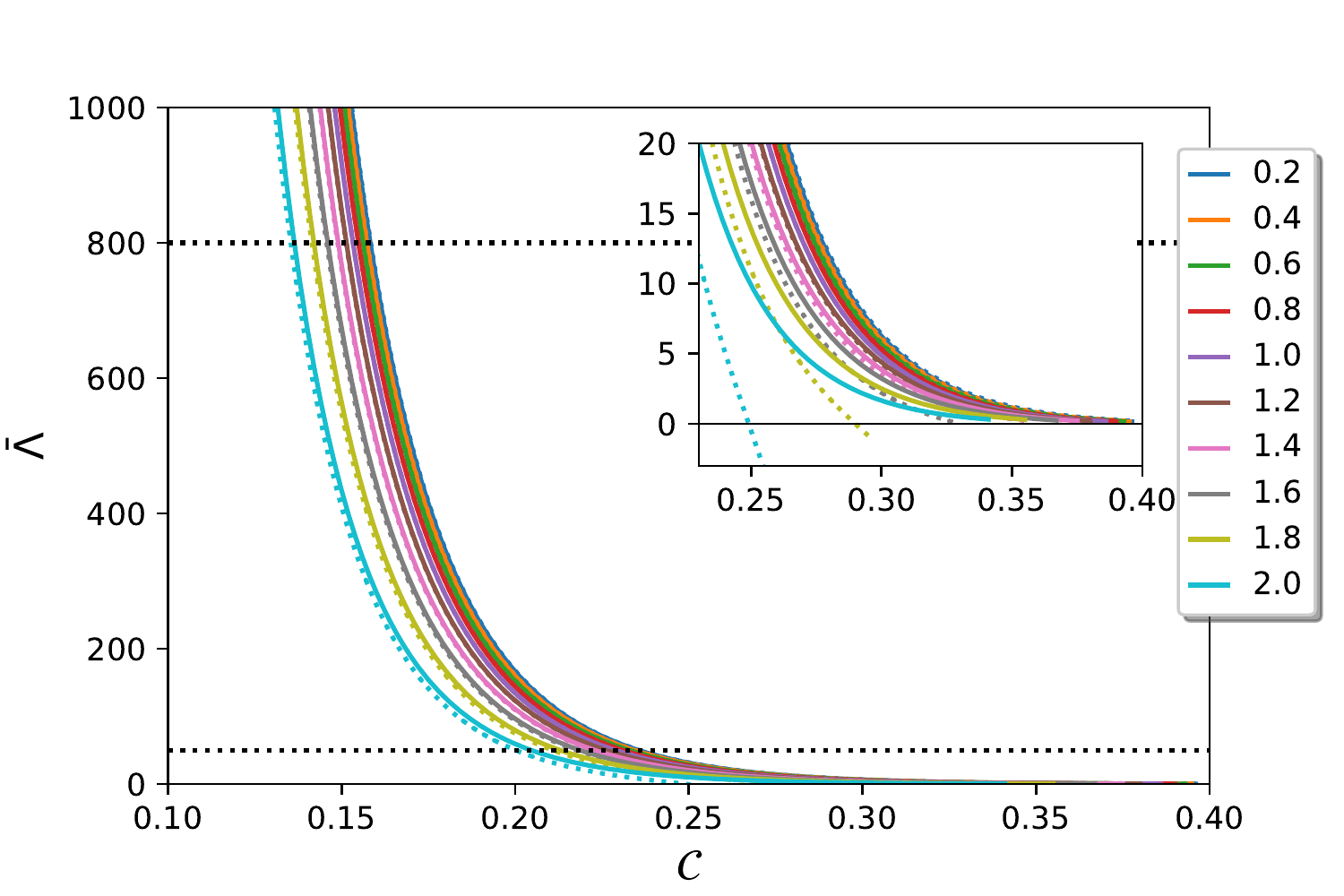}
    \caption{Dimensionless tidal deformability $\bar{\Lambda}$ as a function of the compactness $\com$ for the EMTVII (solid lines) and MTVII (dotted lines) solutions, for $\alpha\in[0,2]$. The black dotted lines indicate the inferred constraints for $\bar{\Lambda}$ from GW170817 \cite{LIGOScientific:2018hze}. Note that $\bar{\Lambda}$ becomes negative for $\alpha>1.6$ and $\com > 0.25$ for the MTVII solution.}
    \label{fig:10}
\end{figure}

In Fig.~\ref{fig:10} we show our results for $\tilde{\Lambda}$, as a function of the compactness, for the EMTVII (solid lines) and MTVII (dotted lines) solutions. The black dotted lines indicate the inferred constraints on $\tilde{\Lambda}$ by GW170817, discussed above. Observe that the curves predicted by the EMTVII model decrease monotonically with $\com$ and approach each other tending to the same value as the compactness approaches the corresponding maximum limit. For the MTVII solution, we observe a similar behavior, at least up to $\alpha=1.4$; however, for higher values of $\alpha$, we observe the curves deviating from each other. For $\alpha>1.6$, there is a fast drop to zero and then $\bar{\Lambda}$ takes negative values near the maximum compactness limit. 

We observe that, for the same compactness, a configuration with low $\alpha$ has a larger $\tilde{\Lambda}$ as compared to one with higher values of $\alpha$. For instance, let us consider a NS with $M=1.4~M_{\odot}$ and $R=11.4~\text{km}$, this implies $\com\simeq 0.182$. Using the results for the EMTVII solution, for $\alpha=0.4$, we have $\tilde{\Lambda}= 305.920$; meanwhile for $\alpha=2$ we have $\tilde{\Lambda}= 117.066$. 

Conversely, the same tidal deformability would be compatible with a NS with certain compactness, depending on the value of $\alpha$. Assuming equal masses, the gravitational wave signature could be only distinguishable by the maximum inspiral frequency before the merger, which would be higher for the configuration with bigger compactness.  

As an example, let us take $\tilde{\Lambda}=50$, which is the fifth percentile reported in \cite{LIGOScientific:2018hze}. Using the results for the EMTVII solution, for $\alpha=0.2$ the corresponding NS would have $\com\sim 0.235$, while for $\alpha=2$ it would have $\com\sim0.205$. This small difference would give a maximum gravitational wave frequency during the inspiral $f_{\text{max}}\sim\sqrt{GM/(\pi^2 R^3)}$ which would be around $18\%$ larger for the NS with $\alpha=0.2$. For an equal mass binary $M_1=M_2=1.4~M_{\odot}$, it implies a change from $4.27$ to $5.24\, \text{kHz}$.  

Finally, we observe that the constraints on $\tilde{\Lambda}$ from GW170817, place important limits to the space of parameters $(\com,\alpha)$ for the EMTVII solution. For instance, Fig.~\ref{fig:10} shows that if one restricts $\alpha\in[0.4,1.4]$, as considered by \cite{Jiang2019}, the compactness is constrained between $\com\sim 0.16$ and $0.23$. Note that this regime of compactness is more restrictive than the one considered by \cite{Jiang2019}, namely, $\com[0.05,0.35]$. On the other hand, if we consider the more general scenario where $\alpha\in[0,2]$, the compactness of a NS is restricted from $\com \sim 0.13$ up to $0.24$.

\section{Conclusions}\label{sec:6}

In this paper, we considered the modified Tolman VII solution, which has been proposed recently to describe the interior structure of realistic NSs~\cite{Jiang2019}. Although the energy density profile of the MTVII model, for certain values of $(\com,\alpha)$, can fit well with the realistic EOS for NSs~\cite{Jiang2019}, we found that the relation for the MTVII pressure shows a region of negative pressure near the surface, for certain values of the parameters $(\com,\alpha)$. As a consequence, we found that the MTVII model predicts negative tidal deformability for certain configurations. To alleviate these shortcomings, we presented an EMTVII solution by numerically solving Einstein's equations for the quartic energy density profile introduced in~\cite{Jiang2019}. 

First of all, we found that our EMTVII solution predicts positive pressure throughout the star, for all the allowed values of the parameters $(\com,\alpha)$. By comparing our results for the EMTVII solution with those for the MTVII model, for the $g_{tt}$ metric component and the radial pressure profile, we found that in the restricted regime of $\alpha\in[0.4,1.4]$ considered in~\cite{Jiang2020}, the fractional errors between the MTVII and EMTVII solutions are relatively low. However, considering a more general parameter space $(\com,\alpha)$, for certain combinations of these parameters, the fractional errors can go up to tens of percent. 

However, the most crucial point is the negative tidal deformability predicted by the MTVII model, for certain combinations of parameters $(\com,\alpha)$. Here we showed that our EMTVII solution predicts a positive tidal Love number, in the whole allowed range for $(\com,\alpha)$. Moreover, the behavior of the tidal Love number, as a function of the compactness, is in much better agreement with previous results found for realistic NSs, as compared to what is predicted by the MTVII solution.  

In Ref.~\cite{Jiang2020} the tidal deformability of the MTVII model was studied in the context of the I-Love-$\com$ relations. Their results differ from ours in the following way. Jiang and Yagi~\cite{Jiang2020} were unable to find an exact solution to Eq.~\eqref{ricatti} for MTVII, so they proposed a series expansion solution in $\com$ for $h(r)$. Using this approach, they presented profiles for $\bar{\Lambda}$, for different values of the parameter $\alpha\in[0.4,1.4]$, up to third order in $\com$. 
\\

\begin{acknowledgments}
This work was supported by the Research Centre for Theoretical Physics and Astrophysics, Institute of Physics at the Silesian University in Opava. We thank John C. Miller for useful discussions. We also thank an anonymous referee who provided valuable comments to improve this paper. 
\end{acknowledgments}

\bibliography{EMTVII}

\end{document}